# Using the structural kinome to systematize kinase drug discovery

Zheng Zhao[1] and Philip E. Bourne[1]*

1. School of Data Science and Department of Biomedical Engineering, University of Virginia, Charlottesville, Virginia 22904, United States of America

*Corresponding author

Email: peb6a@virginia.edu

# Abstract

Kinase-targeted drug design is challenging. It requires designing inhibitors that can bind to specific kinases, when all kinase catalytic domains share a common folding scaffold that binds ATP. Thus, obtaining the desired selectivity, given the whole human kinome, is a fundamental task during early-stage drug discovery. This begins with deciphering the kinase-ligand characteristics, analyzing the structure-activity relationships and prioritizing the desired drug molecules across the whole kinome. Currently, there are more than 300 kinases with released PDB structures, which provides a substantial structural basis to gain these necessary insights. Here, we review *in silico* structure-based methods – notably, a function-site interaction fingerprint approach used in exploring the complete human kinome. *In silico* methods can be explored synergistically with multiple cell-based or protein-based assay platforms such as KINOMEscan. We conclude with new drug discovery opportunities associated with kinase signaling networks and using machine/deep learning techniques broadly referred to as structural biomedical data science.

# 1. Introduction

A kinase is an enzyme that catalyzes the transfer of the ATP γ-phosphate to a specific substrate[1-2]. The human kinome comprises 538 known kinases, and these play an important role in the signal transduction and regulation of cellular functions, such as cell proliferation and necrosis[3-4]. Correspondingly, dysfunctional kinases are associated with a variety of diseased conditions, such as cancer, inflammatory disease, cardiovascular disease, neurodegenerative disease, and metabolic disease[5-6]. Therefore, kinases represent important therapeutic targets to overcome these diseases[7] and have become one of the most potentially impactful target families[8-10]. Since the first kinase-targeted drug, imatinib[11], was approved by the US Food and Drug Administration (FDA) in 2001, a significant breakthrough in kinase drug design for cancer treatment[12], 63 small molecule kinase inhibitors have been approved by the FDA[13-14] as of Feb. 12, 2021. These drugs provide a variety of disease treatments, such as for non-small cell lung cancer (NSCLC)[15], chronic myelogenous leukemia (CML)[16], rheumatoid arthritis[17], breast cancer[18], and acute lymphoblastic leukemia (ALL)[19]. However, in practice, the off-target toxicities and other adverse effects, such as congestive heart failure and cardiogenic shock in some CML patients[20], require the further development of more effective, highly selective inhibitors[3].

Attaining such high selectivity is a daunting task since the inhibitor should bind to a specific primary kinase or select kinases, yet all kinase catalytic domains share a common folding scaffold that binds ATP[21]. To validate selectivity, kinome-scale screening of lead compounds has been attracting more attention[22-23]. Indeed, there are a number of experimental kinome-scale screening methods[22, 24], such as KinaseProfiler[25], KinomeScan[26], and KiNativ[27]. Although kinase profiling technologies are gradually maturing, they are expensive, especially for screening a large compound library against the whole kinome, which remains impractical.

With the availability of an increasing number of kinase structures, virtual structure-based drug screening provides a low-cost and effective way to filter a large compound library and identify the most likely compounds at an early stage of drug screening. Used concurrently with experiential profiling platforms in silico methods provide early-stage kinome-scale drug screening. Based on structural insights, the atom-level binding characteristics of every compound can be revealed and can be used as a guideline for further compound identification and optimization. Given the more than 300 kinases with released PDB structures, subtle differences have been found in the vicinity of the binding site where the adenine base of ATP binds, as well as binding sites away from the ATP binding site, such as in the C lobe of the kinase domain[28-29]. This structural corpora provides insights towards achieving the desired selectivity.

In this chapter we describe the characterization of the whole structure kinome to facilitate drug development. Specifically, we use the function-site fingerprint method to analyze the structural kinome providing systematic insights into kinase drug discovery. With increased knowledge of kinase-driven signaling pathways new kinase targets are continuously being explored for related disease treatment. Looking ahead structural biomedical data science, combining structure-based polypharmacology with machine/deep learning new challenges and opportunities are discussed.

## 2. Kinome-level profiling

Due to the common ATP-binding pocket, the kinase domain was thought to be undruggable prior to the 1990s[30]. With advances in protein- and cell-level experimental techniques and an increase in structure-based knowledge of protein kinases, variation among different kinases became apparent[31]. However, possible specificity requires kinome-scale validation. Moreover, with the increased knowledge of kinase signal pathways, comparison of traditional "one-drug-one-target" models, have been replaced by the acceptance of polypharmacology. Examples include the FDA-approved drug Crizotinib targeting ALK and Met for treating NSCLC, and Cabozantinib targeting VEGFR, MET, RET, FLT1/3/4, AXL, and TIE2 for treating thyroid cancer. Hence, kinome-level profiling is an important step in confirming the selectivity of multi-target drugs.

Multiple commercial platforms provide kinome profiling services with panels ranging from 30 to 715 kinases[1, 32-33] (Table 1).

**Table 1**. Commercial kinase profiling service providers as of Feb. 28, 2021, based on the provider's webpages.

| Providers | Technologies | Kinases | Results | Websites |
|---|---|---|---|---|
| Reaction biology | HotSpot™ $^{33}$PanQinase™ | 715 | $IC_{50}$ | https://www.reactionbiology.com |
| DiscoverRx | KinomeScan | 489 | $Kd/IC_{50}$ | https://www.discoverx.com |
| Thermo Fisher Scientific | Z′-LYTE Adapta | >485 | $IC_{50}$/EC50 | https://www.thermofisher.com |
| Eurofins Discovery | KinaseProfiler™ | >420 | $IC_{50}$ | https://www.eurofinsdiscoveryservices.com |
| Luceome Biotechnologies | KinaseSeeker™ KinaseLite™ | 409 | $IC_{50}$ | https://www.luceome.com |
| ActivX Biosciences | KiNativ™ | >400 | $Kd/IC_{50}$ | https://www.kinativ.com |

Apart from revealing off-targets, profiling inhibitors offers new opportunities for drug discovery[23, 34-39]. Through profiling, the target spectrum reveals the compound's selectivity based on the coverage of kinases it hits, including unexpected off-target interactions, which is a cost-effective way of jumpstarting new kinase drug discovery[40]. For example, Druker et. al. utilized an *in vitro* profiling panel of 30 kinases to establish the selectivity of imatinib in 1996[16]. Later scientists revisited the successful drug using a larger profiling panel and found that imatinib has multiple off-targets in the human kinome. By utilizing the off-target interactions, imatinib can be repurposed for other diseases. Indeed, in 2008 the FDA approved imatinib as an adjuvant treatment to CD117-positive gastrointestinal stromal tumors (GIST) in adult patients[41-42].

## 3. Structural kinome

Profiling the whole human kinome as a routine procedure can validate the selectivity of any given compound by comparing the binding affinities, such as $IC_{50}$ or $Kd$[10, 43]. Subsequently optimizing the compound toward the desired selectivity is the next critical step in early-stage drug discovery. This often begins with deciphering the kinase-ligand characteristics and analyzing the structure-activity relationships, such as confirming which part of the binding sites is nucleophilic/electrophilic, which sub-pocket is hydrophobic, or which amino acids can provide covalent interactions. These atom-level interaction details provide the basic principles by which to modify the functional groups of the given compound. Iteratively combining compound optimization with kinome profiling establishes lead compounds for further testing.

As of Feb. 2021, there are 304 kinases associated with 5208 PDB structures covering all kinase groups, i.e., AGC (276 structures), Atypical (255 structures), CAMK (587 structures), CK1 (82 structures), CMGC (1428 structures), STE (296 structures), TK (1447 structures), TKL (335 structures), and other (493 structures)[44-45]. The 3-D kinase structure corpus provides a basis for structural kinome-based drug discovery. Scientists can not only directly review the compound binding details against the specific target, but they can also compare nuances -similarities and differences - among different kinase targets. For example, in comparing the ATP binding mode, 63 FDA-approved small-molecule kinase drugs can be divided into Type-I, II, III, or IV inhibitors[29, 46]. Similarly, based on the possible existence of a covalent interactions, these kinase drugs can be divided into covalent (irreversible) inhibitors and noncovalent (reversible) inhibitors[47-48] (Table 2). Overall, the desired selectivity is achieved by utilizing every nuance of the different binding sites and accommodating the different sub-pockets of the binding sites among the different kinases[13, 46, 49]. As such, deciphering the whole structural kinome will be very useful in enhancing kinase inhibitor screening, optimization, and prediction. To this end, we introduce the alignment of the binding sites across the structural kinome and describe the characteristics of the aligned binding sites for achieving the desired selectivity.

**Table 2**. 63 FDA-approved kinase small molecule drugs as of Feb. 12, 2021. Columns 2-4 show that the inhibitor types (column 2), covalent interaction modes (column 3), and PDB IDs if the drug-bound structure is available in the PDB (column 4; "-" means the drug-bound PDB structure is unavailable).

| Drug | Type | Mode | PDB ID | Drug | Type | Mode | PDB ID |
|---|---|---|---|---|---|---|---|
| Imatinib | II | Reversible | 1OPJ | Brigatinib | I | Reversible | 5J7H |
| Gefitinib | I | Reversible | 4I22 | Midostaurin | I | Reversible | 4NCT |
| Erlotinib | I | Reversible | 4HJO | Neratinib | I | Irreversible | 2JIV |
| Sorafenib | II | Reversible | 4ASD | Abemaciclib | I | Reversible | 5L2S |
| Sunitinib | I | Reversible | 2Y7J | Copanlisib | I | Reversible | 5G2N |
| Dasatinib | I | Reversible | 3QLG | Acalabrutinib | I | Irreversible | - |
| Lapatinib | I | Reversible | 1XKK | Netarsudil | I | Reversible | - |
| Nilotinib | II | Reversible | 3GP0 | Fostamatinib | I | Reversible | 3FQS |
| Pazopanib | I | Reversible | - | Baricitinib | I | Reversible | 4W9X |
| Vandetanib | I | Reversible | 2IVU | Binimetinib | III | Reversible | 6V2X |
| Crizotinib | I | Reversible | 3ZBF | Encorafenib | I | Reversible | - |
| Vemurafenib | I | Reversible | 3OG7 | Dacomitinib | I | Irreversible | 4I24 |
| Ruxolitinib | I | Reversible | 4U5J | Gilteritinib | I | Reversible | 7AB1 |
| Axitinib | I | Reversible | 4AGC | Larotrectinib | I | Reversible | - |
| Bosutinib | I | Reversible | 4OTW | Lorlatinib | I | Reversible | 5A9U |
| Regorafenib | II | Reversible | - | Entrectinib | I | Reversible | 5FTO |
| Tofacitinib | I | Reversible | 3LXN | Erdafitinib | I | Reversible | 5EW8 |
| Cabozantinib | II | Reversible | - | Fedratinib | I | Reversible | 6VNE |
| Ponatinib | II | Reversible | 4C8B | Pexidartinib | II | Reversible | 4R7H |
| Trametinib | III | Reversible | 7JUR | Upadacitinib | I | Reversible | - |
| Dabrafenib | I | Reversible | 4XV2 | Zanubrutinib | I | Irreversible | 6J6M |
| Afatinib | I | Irreversible | 4G5J | Pemigatinib | I | Rreversible | - |
| Ibrutinib | I | Irreversible | 5P9I | Pralsetinib | I | Rreversible | 7JU5 |
| Ceritinib | I | Reversible | 4MKC | Ripretinib | II | Reversible | 6MOB |
| Idelalisib | I | Reversible | 4XE0 | Selpercatinib | I | Rreversible | 7JU6 |
| Nintedanib | I | Reversible | 3C7Q | Selumetinib | III | Reversible | 4U7Z |
| Palbociclib | I | Reversible | 2EUF | Tucatinib | I | Rreversible | - |
| Lenvatinib | I | Reversible | 3WZD | Avapritinib | I | Rreversible | - |
| Cobimetinib | III | Reversible | 4AN2 | Capmatinib | I | Rreversible | 5EOB |
| Osimertinib | I | Irreversible | 4ZAU | Tepotinib | I | Rreversible | 4R1V |
| Alectinib | I | Reversible | 5XV7 | Trilaciclib | I | Rreversible | - |
| Ribociclib | I | Reversible | 5L2T | | | | |

Zhao et. al. developed a function-site interaction fingerprint (FsIFP) approach to align and delineate the structural human kinome[29]. The FsIFP approach describes protein−ligand interaction characteristics at the functional site using 1D fingerprints[50], which can be compared and contrasted. The approach comprises three steps (Figure 1). First, preparing the structural kinome database. All the released PDB structures can be directly downloaded based on keyword search or the kinase enzyme index[44]. Additionally, there are a few kinase-specific webpages such as UniProt (https://www.uniprot.org/docs/pkinfam), KLIFS[45](https://klifs.net/), and the kinase sarfari https://chembl.gitbook.io/chembl-interface-documentation/legacy-resources#kinase-sarfari). Second, aligning the binding sites. The SMAP software[51], one sequence-independent binding site comparison tool, was applied to compare all of the binding sites. Third, encoding the interaction fingerprint. Given any protein-ligand complex, every involved residue comprising the functional site is encoded as a seven-bit fingerprint using predefined geometric rules[52] for seven types of interaction: (1) van der Waals; (2) aromatic face-to-face; (3) aromatic edge-to-face; (4) hydrogen bond (protein as hydrogen bond donor); (5) hydrogen bond (protein as hydrogen bond acceptor); (6) electrostatic interaction (protein positively charged); and (7) electrostatic interaction (protein negatively charged)[53]. Currently, there are a few open-source tools available, such as IChem[54] and PyPlif[55] that provide these data. The function-site interaction fingerprints are obtained by combining the aligned binding sites with the encoded interaction fingerprints. So far, the FsIFP strategy has been successfully applied to a number of different drug design and discovery projects[15, 56-58].

The FsIFP approach, which examines the specificity among binding sites has been explored to design high-selectivity kinase inhibitors. Beyond the ATP binding pocket, there are other binding sites to be validated[59], such as hydrophobic segment, allosteric segment, DFP motif area, and G-rich-loop region (Figure 1a). Corresponding to these binding regions, inhibitors are classified as Type-I, Type-II, and Type-III.

Type-I kinase inhibitors mainly bind to the ATP-binding site in the "DFG-in" conformation. To obtain stronger binding affinity and greater selectivity than ATP, besides occupying the ATP-binding space, Type-I inhibitors extend into different proximal regions, specifically referred to as the front pocket region, the hydrophobic pocket region, the DFG motif, or the G-rich-loop region[13, 59]. For example, Gefitinib is one Type-I drug for the treatment of non-small cell lung cancer (NSCLC)[60]. Its quinazoline scaffold forms hydrogen bonds with the hinge region like the adenine moiety of ATP (Figure 1b). More importantly, the 3-chloro-4-fluorophenyl fragment of Gefitinib extends into the hydrophobic pocket, and the morpholine derivative binds at the front pocket and forms polar interactions with residues Cys797 and Asp800 (Fig. 2a)[61]. In contrast, Type-II kinase inhibitor typically bind in the "DFG-out" conformation. Type-II kinase inhibitors extend into the allosteric pocket region beyond the ATP binding pocket. For example, Imatinib is a Type-II inhibitor to treat positive acute lymphoblastic leukemia (Ph+ ALL) in children. Like Type-I inhibitors, there is a scaffold fragment (Figure 1c) occupying the space where the adenine moity of ATP binds. At the same time, the 4-(4-Methyl-piperazin-1-ylmethyl) benzamide extends into the allosteric pocket. Type-III inhibitors occupy the allosteric pocket (Figure 1d), which is not so well conserved that attractive for designing non-ATP competitive kinase inhibitors. Fingerprints established through aligned structures provide detailed information on the binding sites occupied by Type- I, II, and III kinase inhibitors.

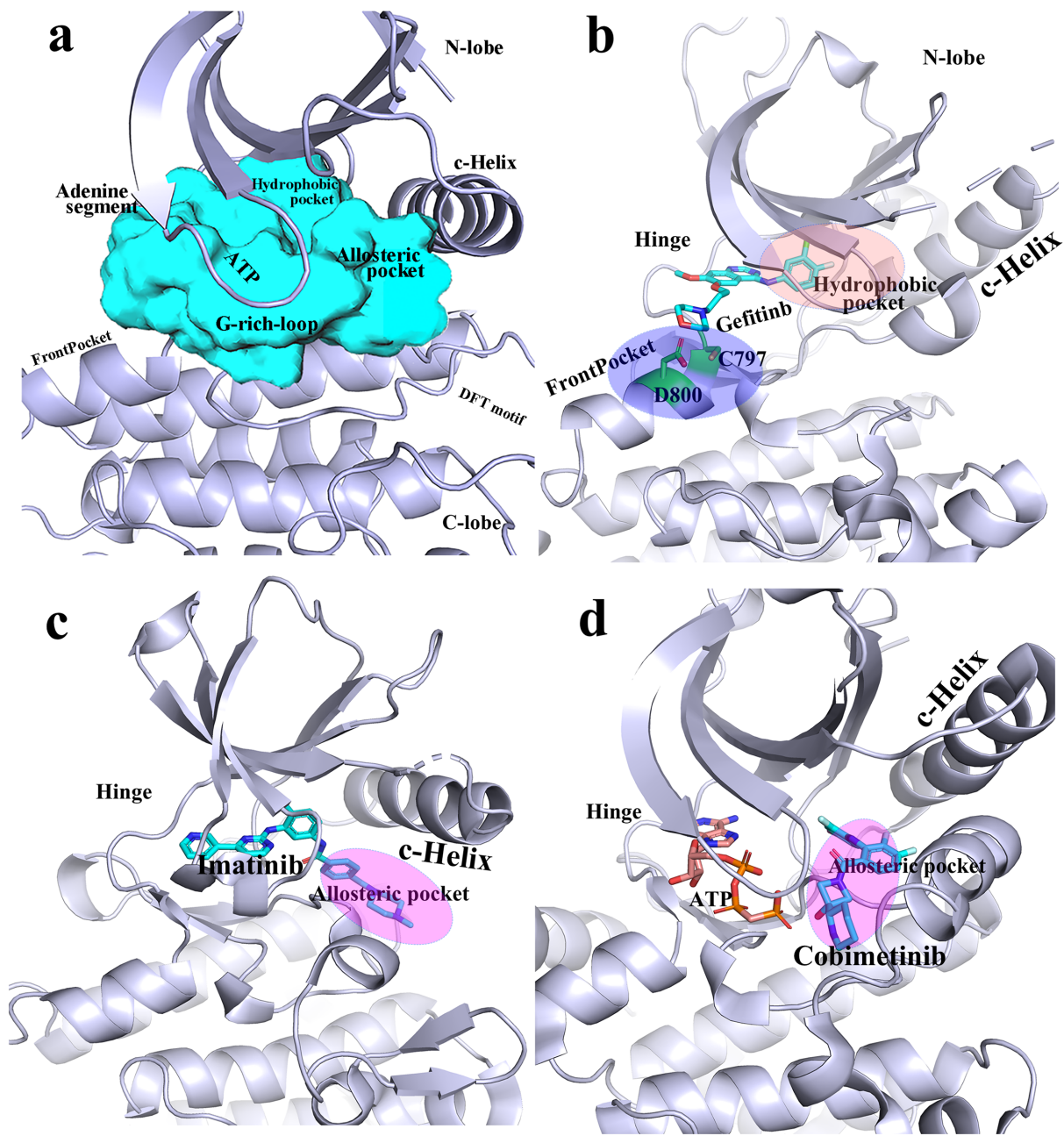

**Figure 1.** (a) Kinase binding site surrounding the ATP binding cavity (PDB ID: 1m17 as the template). (b). The occupied binding pocket of a Type-I kinase inhibitor (PDB ID: 4i22). (c) The binding characteristics of a Type-II kinase inhibitor (PDB ID: 2hyy); (d) The binding characteristics of a Type-III kinase inhibitor (PDB ID: 4an2).

The other noteworthy aspect is the difference in amino acids at the same spatial position among the aligned binding sites, which is useful in obtaining the desired selectivity. Typically, structural kinome-guided studies have shown that there are a number of cysteine residues distributed around the binding sites[1, 21, 62-63]. In Zhang et. al.'s review, they identified over 200 kinases bearing at least one cysteine in and around the ATP binding pocket, thus highlighting the broad structural basis to improve the selectivity and binding affinity by covalently utilizing these non-catalytic cysteines[21]. Currently, 7 covalent kinase inhibitors have been approved (Table 2). This class of inhibitors not only bind to the ATP binding pocket, but also nearby cysteines to form covalent interaction; Afatinib (Figure 2a) being one example through binding to Cys797. Similarly, the other 6 inhibitors all form covalent bonds with the corresponding cysteines located in the front pocket. Like Afatinib, Osimertinib, Dacomitinib, and Neratinib all have covalent interaction with Cys797 in the EGFR crystal structure (Figure 2a). Ibrutinib, Acalabrutinib, and Zanubrutinib all form covalent interactions with Cys481, targeting BTK. In their study, Zhao et. al.[63] identified 63 different amino acid locations bearing accessible cysteines through surveying the whole structural kinome (Figure 2b), speaking to the broad applicability of designing covalent kinase inhibitors.

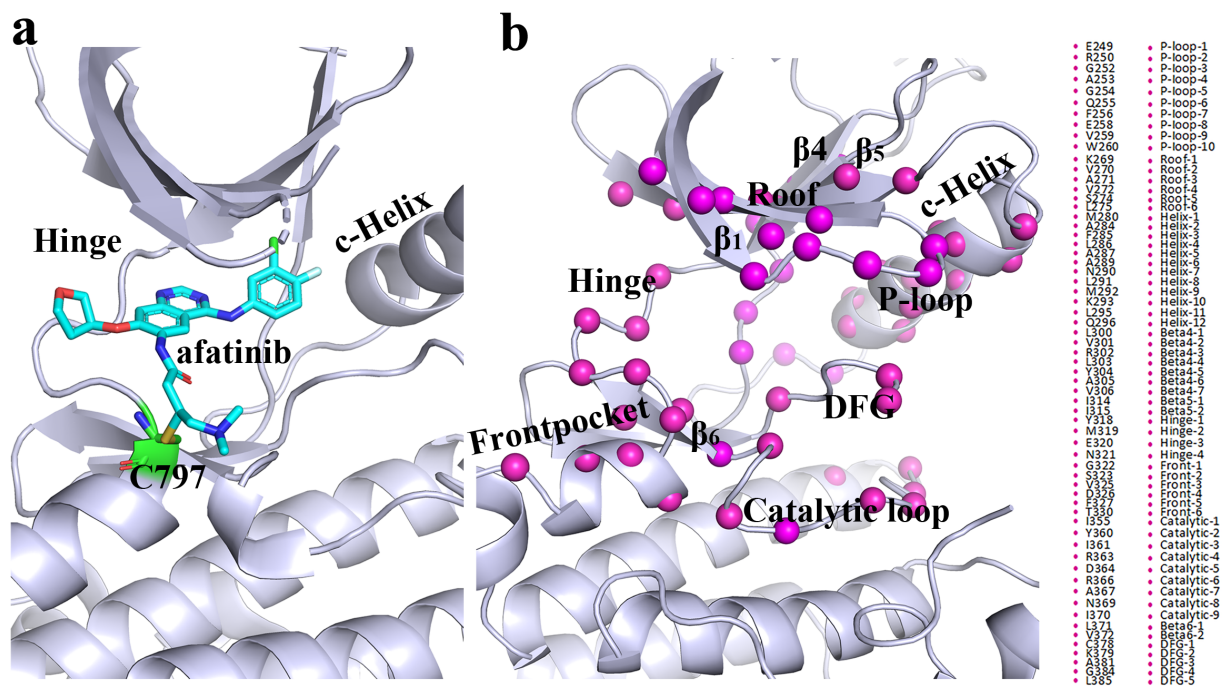

**Figure 2.** (a) Covalent binding mode of Afatinib (PDB id: 4g5j). (b). 63 positions bearing the accessible cysteines (PDB id: 3byu). The legend shows the amino acids and the corresponding spatial positions (purple balls).

## 4. Challenges and Opportunities

Since the launch of the first kinase drug, Imatinib in 2001, kinase targeted drug discovery has been on a fast track. In the last six years, an average of eight small molecule kinase drugs have been approved per year. This tremendous success benefits patients, but also highlights our ability to achieve drug discovery outcomes[9, 64]. However, challenges still remain in the development of efficient, non-toxic kinase-targeted drugs[3].

Clinical adverse effects are one major challenge. For example, kinase drugs affect the digestive system and cause nausea, vomiting, and/or diarrhoea[65]. Further, most kinase inhibitors cause serious adverse effects, such as different degrees of cytopenia[66]. These side effects typically result from off-targets effects. To avoid such side effects, a highly selective drug is desired. Alternatively, adverse effect can be due to on-target toxicities involving the intrinsic mechanisms of the drugs[67]. At this point in the evolution of small molecule kinase drugs novel compound scaffolds are needed. to reduce adverse effect as much as possible. Scaffolds that are available for early-stage screening. The increasing availability of panels of phenotypic assays may provide one strategy to profile selectivity by combining virtual structure-based kinome screening, which can filter a huge compound library into a highly focused kinase library[68].

Another challenge is acquired drug resistance[58, 69]. In clinical practice, kinase-targeted drugs are frequently subject to drug resistance, which has become a primary vulnerability in targeted cancer therapy. The first difficulty is exploring resistance mechanisms due to the diversity of specific drug-binding mechanisms. For example, drug resistance of Erlotinib, which is one FDA-approved kinase drug used to treat patients with EGFR-overexpression induced NSCLC, is caused by the gatekeeper T790M mutation, which increases the binding affinity of ATP to the EGFR kinase[70]. In another example, Crizotinib was often found to be ineffective in the majority of patients after 1−2 years' treatment against ALK-positive NSCLC due to the acquired ALK L1196M mutation, which decreased the binding affinity of Crizotinib[71].

Nevertheless, these challenges also provide unique opportunities to develop new approaches and applications. Currently, in vitro and/or in vivo kinome-scale and proteome-scale profiling methodologies have been merged into the drug design pipeline, which potentially provides a thorough understanding of targets and selectivity of kinase inhibitors. Combined with the diseases' signal pathway, the target spectrum can be further applied to "one-drug-multiple-target" drug design. For example, Midostaurin is a multi-target kinase drug[72] used to treat adult patients with newly diagnosed FLT3-mutated acute myeloid leukemia (AML). For a "multiple-drug-multiple-target" combination therapy strategy, Capmatinib (a MET inhibitor) and Gefitinib (an EGFR inhibitor) had been approved to treat patients with EGFR-mutated-MET-dysregulated – in particular, MET-amplified - NSCLC[73]. We can expect profiling methodologies will be further developed to cover the whole kinome and even the proteome.

**Table 3.** Kinase-inhibitor interaction activity data resource.

| Provider | Technology | Coverage | Resource website |
|---|---|---|---|
| Reaction biology | HotSpot™ | 300 Kinases × 178 Inhibitors | www.guidetopharmacology.org |
| ChEMBL Kinase SARfari | Collected from academic publications | ~530,000 data points | www.ebi.ac.uk/chembl |
| Eurofins Discovery | KinaseProfiler™ | 234 Kinases × 158 Inhibitors | www.guidetopharmacology.org |
| DiscoverRx | KinomeScan | ~440 Kinases × 182 Inhibitors | lincs.hms.harvard.edu |
| ActivX Biosciences | KiNativ™ | (194 to 316) Kinases × 30 Inhibitors | lincs.hms.harvard.edu |

In virtual drug design and discovery, the incorporation of machine/deep learning and structural biomedical data science are advancing compound screening, target validation, and selectivity improvement[74]. Currently, data science has become one of the fastest-growing disciplines and deep learning has been applied to drug synthesis, design, and prediction[75-77]. Moreover, there are a large number of kinase assay databases available. For example, a database from Merck, KGaA, with over 1.0 million data points (i.e., 4,712 compounds × 220 kinases). Merget et. al used it to train one virtual profiling assay model to support virtual screening, compound repurposing, and the detection of potential off-targets[37]. Here, we collate the free databases of available kinase-inhibitor activity (**Table 3**). It is worth noting that the ChEMBL Kinase SARfari database, which contains ~54,000 compounds, ~980 kinases targets, and the corresponding approximately 530K structure-activity data points[78], has been used to predict kinome-wide profiling of small molecules[79-81].

Taken together, data-driven methods and applications will further experimental protocols and facilitate the drug discovery processes.